\pdfoutput=1

\documentclass[letterpaper, 10 pt, conference]{ieeeconf}  

\IEEEoverridecommandlockouts                              
\title{\LARGE \bf
A Complete Review of Controlling the FDR in a Multiple Comparison Problem Framework- The Benjamini-Hochberg Algorithm
}

\author{$Anish Acharya^1$
\thanks{$^{1}$A. Acharya is with Faculty of Electrical Engineering, University of California, Irvine
        {\tt\small anisha@uci.edu}}
}

\begin{document}

\maketitle
\thispagestyle{empty}
\pagestyle{empty}


\section{Abstract}
The paper titled 'Controlling the False Discovery Rate-a practical and powerful Approach to multiple Testing' by benjamini et. al.[1] proposes a new framework of controlling the False Discovery Rate in a Multiple Hypothesis testing problem. It has been claimed that the procedure proposed in the paper results in a substantial gain in power more applicable in case of problems which call for False discovery rate (FDR) control rather than Familywise Error Rate (FWER). The proposed method uses a simple Bonferroni type procedure for FDR control.

\section{Background and Summary of Technical contribution}
In the field of Decsion and Inference making Hypothesis testing plays a crucial role. A parameter can be estimated looking at the obtained samples, either by a single number (point estimation) or by an entire interval of plausible values (confidence interval). However, in most cases the object is not to estimate a parameter but to decide which of the contradictory claims about the parameter is correct or more likely. The theory of decision making from a set of contradictory claims is known as Hypothesis testing. 

\subsection{Single Hypothesis Test}

In case of simple Hypothesis testing problem, there are two contradictory hypothesis. The established one or the Null hypothesis and one that contradicts the established theory, the alternative hypothesis. The first is to formulate a test statistics which is a function of random samples. Now, given the sample instant one gets an outcome of the test statistic. the distribution of the test statistic under null hypothesis is known and is based on the assumptioins made in the probabilistic model. Based on the outcome of the test statistic a desicion is made under a pre-decided decision rule. 
\\In decision making two types of error might arise- 
\begin{itemize}
\item Type-I error- 
\\ Wrongly rejecting the null Hypothesis is termed as the Type-I error(false positive).
i.e. $H_1\mid\theta\in\theta_0$
\item Type-II error-
\\ Wrongly accepting null hypothesis is termed as Tpe-II error or False negative. i.e. $H_0\mid\theta\in\theta_1$ 

\end{itemize}

Now one would like to keep both these error as small as possible. however, unfortunately these errors are inversely related to each other i.e. if one is decreased the other is increased. So there has to be a trade off between these two errors. Here for obvious reasons Type-I errors are delicately handeled as they are more important and should be taken care of. So, the basic idea is to fix a significance level i.e. the tolerance level of Type-I error and minimize the type-II error. The probability of type-I error called the significance level or level of test denoted by $sup_{\theta\in\theta_0}Pr(H_1\mid\theta\in\theta_0)$. However, in applications this level of $\alpha$ is somewhat arbitrary and. Thus, in practice one might have different conclusions for different values of $\alpha$. Thus, the concept of p-value or attained significance comes into play. It is defined as the probability of the given data under the null hypothesis. In simpler words, it signifies if the Null Hypothesis was true then how likely we would observe the sample data. In other words it is the smallest level attained significance level, i.e. the smallest level of significance for which the null hypothesis would be rejected. The smaller the p-value is the stronger the evidence behind rejecting the null hypothesis. An alternative approach to this p-value would be come up with a rejection region $\Gamma$based on the pre-defined significance level and if the outcome of the test statistic $t\in\Gamma$ then we reject the null hypothesis else we keep the null hypothesis. Now,if one chooses high threshold then it is more likely to attain higher power but prone to false positive there has to be a trade-off between high power and false negative.  

\subsection{Multiple Hypothesis Testing}
Multiple hypothesis Testing is quite common in practice and often encountered. An example would be: Suppose one has several forecasting strategies to compare with the benchmark and wishes to identify which works better than the existing benchmark.

In order to test several hypothesis simultaneously the first idea would be to test each hypothesis separately w.r.t some significance level $\alpha$. however keeping in mind the multiplicity this might not be a good idea to pursue. For example, if we have a significance level 0.05 one would expect 5 Type-I errors in 100 samples even if all the test are not significant. Now, if we have 20 hypothesis to test then it turns out to be $1-(1-0.05)^{20}=0.64 $ considering the hypothesis to be independent with each other and this is quite unacceptable that even if all the hypothesis are insignificant still the probability of making a Type-I error is 64 out of 100 and as a matter of fact mostly in practical cases the number of hypothesis to be tested is in general much larger than 20 and unfortunately due to the chance this error rate keeps on going up. The idea is to come up with an adjustment of $\alpha$ so that the error rate remains below the desired significance level. To deal with this multiplicity problem Classical multiple Comparison invoke the Familywise error rate(FWER) defined as the probability of one or more False Alarms. 

FWER=Pr($\cup_{i=1}^{m}$Pr($H_{1i}$when $H_{0i}$is true)
\\Bonferroni Inewuality or the Union Bound suggests 

Pr($\cup_{i=1}^{m}A_i$)$\leq$ $\sum_{i=1}^mPr(A_i)$
\\Thus, as a natural extension of this well known Union Bound it can be concluded that if each hypothesis is tested at the level $\alpha$ then FWER$\leq m\alpha$ Thus to get around this problem and ensuring FWER $\leq\alpha$ for the family of m tests the Bonferroni Correction Suggests to set the significance level to $\frac{\alpha}{m}$ thus the threshold is now selected as $\gamma=Q^{-1}(\frac{\alpha}{m})$ instead of $\gamma=Q^{-1}(\alpha)$ and thus FWER$\leq\alpha$ As can be seen that this Bonferroni correction scheme is qquite conservative. Here though we benifit from the independent assumptions but in practice this is hardly the case and keeping in mind the correlation structure of the tests, it can be concluded that this Bonferroni correction could be extremely conservative as it leads to a high number of false negativeness. The Per Comparison Error rate (PCER) approach that almost ignores the multiplicity problem alltogether is defined as PCER=E($\frac{V}{m}$); where V is the Number of type-I error s made or the number of false alarms and m is the total number of hypothesis being tested.Testing each hypothesis at the significance level $\alpha$ ensures PCER$\leq\alpha$

\subsection{Summary of Contribution of this paper}
The authors pointed out some of the practical disadvantages of FWER and PCER and the necessity of an alternative and more powerful procedure as follows-
\begin{itemize}
\item Most of the methods of FWER controlling MCPs assume multiple-treatment type problem and thus concerns multivariate normal distribution which in practice is not always the case.
\item FWER tends to have lower power than the PCER at levels conventional to single-comparison problems. 
\item FWER as can be seen from the definition might reject when atleast one is significant. This is likely when several methods are compared with a baseline and the best will be selected. However, in many cases viz. in various treatment group/control group various aspects of the treatment are compared and it can't be concluded that the existing treatment is errornous even if some of the null hypothesis are rejected. Thus, in these cases FWER based error control strategies are not suitable. 
\end{itemize}

Thus the authors have proposed a alternative approach to deal with the MCP frameworks. For that a new type of error measure has been invoked and termed as "False Discovery Ratio(FDR)".Suppose m be total number of Hypothesis being tested,$m_0$be the number of True Null Hypothesis. Out of these $m_0$true null hypothesis U are declared non-significant and V are declared significant (False Alarm). And out of the $m-m_0$ alternative Hypothesis S are rightly identified as significant and T are misclassified as non-significant(Type-II error/Flase Negative). However, U,S,V,T are unobservable random variables and the statistician observes only total number of Significant outcomes R=V+S and total declared insignificant m-R=U+T.

The authors define FDR as the proportion of errors committed by falsely rejecting null hypothesis which can be transletted into the term Q=$\frac{V}{V+S}$. Q=0 when V+S=0 due to the fact in this case no error can be committed. However, Q is an unobserved quantity as V,S are both unobservable. Thus, FDR is defined as-

FDR=$Q_e=E(Q)=E(\frac{V}{V+S})=E(\frac{V}{R})$

Two important aspects are pointed out here -
\begin{itemize}
\item When all the Null Hypothesis are true then S=0 so, FDR=E(Q)=Pr(V$\geq$1)

which is exactly equal to FWER thus, in it has been claimed that controlling FDR automatically controls FWER. 
\item When all hypothesis are not null (more common case) then FDR$\leq$FWER as V$\leq$R thus,Pr(V$\geq1)\leq Q_e$ Thus it has been shown that the FDR is often more powerful than FWER. 
\end{itemize}

The algorithm proposed by the authors in support of their claims can be in short broken in three step concisely to Control FDR at level $\alpha$ 
\begin{itemize}
\item Test the Hypothesis $H_1,H_2...H_m$ based on the corresponding p-values $p_1,p_2,....p_m$
\item Order the p-values $p_1\leq p_2\leq p_3\leq ...\leq p_m$
\item Rank the Hypothesis according to their corresponding p-values. 
\item Find the test with highest rank j for which the p value $p_j\leq \frac{j}{m}\alpha$
\item Declare the top j tests 1,2,....j with $p_j\leq \frac{j}{m}\alpha$ as significant. 
\end{itemize}
the authors have reported this as a theorem and provided a comprehensive proof to claim that this method will always lead to controlled FDR at $\alpha$

The paper uses an example of treatment procedures pertaining to cardiac patients. With the help of a few example it shows that in case of inference problem where the comparison is not between different methods rather the comparison is between different features the BH FDR method finds out significant results when the FWER based method fail to do so. The author's also put forward a different way to formulate their proposed algorithm. 
It has been claimed that the proposed algorithm can be formulated as a constraint optimization problem. it has been suggested that it is same as to choose $\theta$ that maximizes the number of rejections, $r(\theta)$ at the level subject to the constraint $\frac{\theta m}{r(\theta)}\leq \alpha$ The author's also claim the method is more powerful as compared to classical FWER based methods with a set of experiments where both the algorithms were tested and the proposed algorithm resulted in much better results.  

\section{Review Comments}
This paper by Benjamini and Hochberg et.al. seems to open a new dimension in the study of multiple comparison test. In order to undestand the effect or the real contribution of the paper in the cotext of multiple comparison test we should think about the preceding works and only then one would be able to evaluate the novelty of the paper. 

When one conducts a multiple comparison test with a fixed level of significance $\alpha$ for all the experiments and carry out the comparison it is pretty straightforward to notice that the Probability of comminting a Type-I error hence the probability of getting significant result goes up just by chance due to the law of probability. As such if there are n experiments being compared then the \\Pr(atleast one type-I error)=1-$(1-\alpha)^{n}$ \\which is not desired. Thus there needed to be a modificatio in $\alpha$ to control this FWER. Thus, looking at this the first thing that comes to the the mind is to reduce the $\alpha$ but in order to do so it would give rise to a reduction in power. Thus, pretty much as one would try straightaway is to reduce the FWER without much loss in power thus, a trade off between them is desired. Sidak et.al. came up with the first form of correction known as Dunn-sidak procedure [2]. They proposed a modified value of the tolerance level. According to there proposal if one chooses the significance level to be $(1-(1-\alpha_e)^{1/n}))$ for each experiment that will ensure the FWER to be controlled at $\alpha_e$. Later Bonferroni came up with a correction which was just the first order approximation of the infinite sequence of proposed Dann-sidak method and the popular Bonferroni correction turns out to be using $\frac{\alpha_e}{n}$ for each of the comparison tolerance level in order to achieve FWER controlled at tolerance level $\alpha_e$. However, both these Boferroni and Sidak Procedures seem to be very conservative thus less powerful. The next significant work was done by Holm S. et. al. [3]. They proposed to rank the p values in ascending order and test the i th p-value $p_i$ against the level $\frac{\alpha}{n-i+1}$ to keep th FWER controlled at $\alpha$. This method turned out to be more powerful the existing Bonferroni Method. Now, the authors of this paper have proposed a modification of the Holm's method and also proposed a new controlling error rate called 'False Discovery rate (FDR)' and arranges p-values in increasing order and tests i th p vaue $p_i$ with $\frac{iq^*}{m}$ where m is the number of hypothesis being tested.Then rejecting all $H_i$ $\forall i\in 1(1)k$ such that $p_i \leq \frac{iq^*}{m}$ and this will ensure that FDR is controlled at the level $q^*$. It also proves that controlling FDR automatically controls the FWER.
\\ Thus, judging from the point of novelty, it seems like an extension to the Holm's method to control FWER and follows the similar procedure. However the introduction of the FDR as an error measurment criterion is very novel and significant as it automatically takes care of FWER control and also additionaly less conservative and thus leading to more powerful detection. 
\\In Section 4.2 the author's have reported relative study between the current method and the existing Bonferroni and Holm-hochberg method (1988) and they have reported that this proposed method leads to Uniformly most powerful among them and its performance is significantly better than the other methods as the number of hypothesis tested are increased. However some technical and non-technical advice to improve the quality and readability of the paper-
\begin{itemize}
\item The quality of the diagrams should be improved for better readability (300 dpi). 
\item It proves significant improvement as compared to the existing procedure however, it doesn't mention if this is the best one can achieve. As in, it would be nice to look into if there is a bound on the power that can be achieved while keeping the FDR level controlled at the desired level. 
\item More simulation study should be added to investigate how does the method performs at different FDR levels with respect to sparsity. 
\item more rigorous analysis should have been made in order to observe if the trend really persists as the number of hypothesis increases to higher levels like m=500 or 1000. It would be nice to look into a long term Power vs. Number of hypothesis analysis and see how powerful the method remains at that high number of hypothesis.

\end{itemize}
Overall, the paper provides nice intuitive development of the Multiple Comparison scheme and goes beyond the preceding work. The idea of FDR introduced here looks very novel and thus opens up a new line of research in this field. Further investigation and extension over this method should follow from this paper.

\end{document}